\documentclass[a4paper,conference]{IEEEtran}
\usepackage{amsmath,amsfonts}
\usepackage{color}
\usepackage{array}
\usepackage{algorithmic}
\usepackage[linesnumbered,ruled,vlined]{algorithm2e}
\usepackage[caption=false,font=normalsize,labelfont=sf,textfont=sf]{subfig}
\usepackage{textcomp}
\usepackage{stfloats}
\usepackage{url}
\usepackage{times}
\usepackage{verbatim}
\usepackage{graphicx}
\usepackage{longtable}
\usepackage[utf8]{inputenc}
\usepackage[T1]{fontenc}
\usepackage{lipsum}
\usepackage{multirow}
\usepackage{mathtools}
\begin{document}
\makeatletter
\newcommand{\nosemic}{\renewcommand{\@endalgocfline}{\relax}}
\newcommand{\dosemic}{\renewcommand{\@endalgocfline}{\algocf@endline}}
\newcommand{\pushline}{\Indp}
\newcommand{\popline}{\Indm\dosemic}
\let\oldnl\nl
\newcommand{\nonl}{\renewcommand{\nl}{\let\nl\oldnl}}
\makeatother
\title{Sequential Multiuser Scheduling and Power Allocation for Cell-Free Multiple-Antenna Networks}

\author{Saeed Mashdour$^{\star}$, Rodrigo C. de Lamare $^{\star,\dagger}$, Anke Schmeink$^{\star\star}$ and João P. S. H. Lima $^{\ddagger}$ \\  $^{\star}$ Centre for Telecommunications Studies, Pontifical Catholic University of Rio de Janeiro, Brazil \\
$^{\dagger}$ Department of Electronic Engineering, University of York, United Kingdom \\
$^{\star\star}$ Chair of Information Theory and Data Analytics,
RWTH Aachen University, 52062 Aachen, Germany\\
$^{\ddagger}$ CPQD, Campinas, Brazil \\
smashdour@gmail.com, delamare@cetuc.puc-rio.br, schmeink@inda.rwth-aachen.de, jsales@cpqd.com.br  \vspace{-2em}\\
\thanks{This work was supported by CNPQ, CPQD, FAPERJ and Funttel/Finep - Grant No. 01.20.0179.00.}}

\maketitle

\begin{abstract}
Resource allocation is a fundamental task in cell-free (CF) massive multi-input multi-output (MIMO) systems, which can effectively improve the network performance. In this paper, we study the downlink of CF MIMO networks with network clustering and linear precoding, and develop a sequential multiuser scheduling and power allocation scheme. In particular, we present a multiuser scheduling algorithm based on greedy techniques and a gradient ascent {(GA)} power allocation algorithm for sum-rate maximization when imperfect channel state information (CSI) is considered. Numerical results show the superiority of the proposed sequential scheduling and power allocation scheme and algorithms to existing approaches while reducing the computational complexity and the signaling load.\\
\end{abstract}
\vspace{-1mm}
\section{Introduction}
\label{sec:intro}

Cell-free (CF) massive MIMO networks introduced in \cite{ngo2017cell} are distributed massive MIMO networks \cite{mmimo,wence} that include several access points (APs) over a geographic area to serve user equipments (UEs) in the same time and frequency resources, which provides uniform performance across users and improves coverage. Since CF networks {in which all UEs are served by all APs,} need to process all channels and signals in a central processing unit at the same time, they result in a huge burden to the processors and a substantial increase in costs. Therefore, it is necessary to use clustering techniques, which include network-centric and user-centric approaches \cite{bjornson2020scalable} to reduce signaling and computational costs.

Resource allocation including power allocation and multiuser scheduling are key tasks for CF networks that can improve the system performance and have attracted a lot of attention in the literature \cite{cheng2016optimal, nayebi2016performance, denis2021improving, palhares2021robust, gong2022dynamic, riera2019trade,tds_cl,tds,jpais,jpba,siprec,baplnc,jlrpa,jlrpa2,rmmseprec,mlrs,srmax,rapa}. Multiuser scheduling can reduce multiuser interference in CF networks, improving the system performance. In addition, if the number of receive antennas is larger than those of transmit antennas, it is impossible to support all the receivers which makes it necessary to employ multiuser scheduling. In this context, power allocation also leads to significant performance improvement in CF massive MIMO networks. In \cite{ammar2021downlink}, the problem of multiuser scheduling, power allocation and beamforming in a user-centric
cell-free MIMO wireless system has been solved by maximization of a weighted sum-rate (WSR) problem. The joint optimization of UE scheduling, power allocation and pilot length is investigated in \cite{ming2022downlink} and the minimum ergodic user rate in the downlink transmission is maximized. In \cite{hamdi2020power}, the energy efficiency of a user-centric CF massive MIMO system is enhanced by solving a total grid power consumption minimization problem with a joint AP selection and user scheduling algorithm.

In this paper, we consider linear minimum mean square error (MMSE) and zero forcing (ZF) precoders and investigate the downlink of clustered CF massive MIMO networks with multiuser scheduling and power allocation. In particular, we develop a sequential multiuser scheduling  and power allocation (SMSPA) scheme based on enhanced greedy and GA techniques to maximize the sum-rate. The proposed enhanced subset greedy (ESG) technique approaches the performance of the optimal exhaustive search method while significant computational cost can be saved. The proposed GA algorithm maximizes the sum-rate and is performed after scheduling the desired UEs set. Simulations show that the proposed SMSPA scheme, ESG and GA algorithms outperform competing approaches.

{\it Notation}: Throughout the paper, $\left \| . \right \|_{F}$ denotes the Frobenius norm, $\mathbf{I}_{N}$ denotes the $N\times N$ identity matrix, the complex normal distribution is represented by $\mathcal{CN}\left ( .,. \right )$, superscripts $^{T}$, $^{\ast}$, and $^{H}$ denote transpose,
complex conjugate and hermitian operations respectively, $\mathcal{A}\cup \mathcal{B} $ is union of sets $\mathcal{A}$ and $\mathcal{B}$, and $\mathcal{A}\setminus \mathcal{B} $ shows exclusion of set $\mathcal{B}$ from set $\mathcal{A}$.
\vspace{-2mm}

\section{System model}
\label{System model}

{The downlink of} a CF massive MIMO network is considered where $K$ uniformly distributed single-antenna UEs are supported by $M$ single-antenna APs. Then, the CF network is clustered by dividing the network area into $C$ non-overlapping clusters so that the cluster $c$ includes $K_c$ uniformly distributed single-antenna UEs and $M_c$ single-antenna APs. We assume that the number of UEs is much larger than the number of APs so that $K>>M$ for the CF network and $K_c>>M_c$ for cluster $c$ of the clustered CF network,{ which requires the scheduling of a subset of $n_c \leq M_c$ UEs per cluster. }
\vspace{-2mm}
\subsection{Cell-free massive MIMO network and the clustered context}
In the CF network, the channel coefficient between AP $m$ and UE $u$ is shown by $g_{m,u}=\sqrt{\beta _{m,u}}h_{m,u}$, \cite{ngo2017cell}, including the large scale fading $\beta _{m,u}$ and the small-scale fading $ h_{m,u}\sim \mathcal{CN}\left ( 0,1 \right )$ defined as independent and identically distributed (i.i.d.) random variables (RVs) constant during a coherence interval and independent over different coherence intervals. After scheduling $n \leq M$ out of $K$ UEs, the received signal is given by
\begin{equation} \label{eq:CF-sig}
\begin{split}
\mathbf{y} & = \sqrt{\rho _{f}}\mathbf{G}^T\mathbf{P}\mathbf{x}+\mathbf{w} \\
 & = \sqrt{\rho _{f}}\hat{\mathbf{G}}^{T}\mathbf{P}\mathbf{x}+\sqrt{\rho _{f}}\tilde{\mathbf{G}}^{T}\mathbf{P}\mathbf{x}+\mathbf{w}
\end{split}
\end{equation}
where $\rho _{f}$ is the maximum transmitted power of each antenna, $\mathbf{G}=\hat{\mathbf{G}}+\tilde{\mathbf{G}}$ is the ${M\times n}$ channel matrix, in which $\hat{\mathbf{G}}$ is the channel estimate, $\tilde{\mathbf{G}}$ is the estimation error that models the CSI imperfection and $\left [ \mathbf{G} \right ]_{m,u}=g_{m,u}$, $m\in \left \{ 1,\ldots ,M \right \}$, $u\in \left \{ 1,\ldots ,n \right \}$, $\mathbf{P}$ is the ${M\times K}$ linear precoder matrix such as MMSE or ZF, $\mathbf{x}=\left [ x_{1},\ldots ,x_{n} \right ]^{T}$ is the zero mean symbol vector with mutually independent elements and independent of the channel coefficients  $\mathbf{x}\sim \mathcal{CN}\left ( \mathbf{0},\mathbf{I}_{n} \right )$, and $\mathbf{w}=\left [ w_{1},\cdots,w_{n}  \right ]^{T}$ is the additive noise vector with $\mathbf{w}\sim \mathcal{CN}\left ( 0,\sigma_{w}^{2}\mathbf{I}_{n} \right )$ and statistically independent of the signal vector. We remark that the design of precoders for this system can rely on several strategies \cite{gbd,mbthp,bbprec,lrrbd,wlbd,memd,baplnc,badstbc,rmbthp,rsrbd,rsthp,rscf,simccdma,cfrmmse,cpm,cqabd,noma}.

Assuming Gaussian signaling, the sum-rate of the CF system is given by
\begin{equation}\label{eq:RCF}
    R_{CF}=\log_{2}\left (  \det\left [\mathbf{R}+\mathbf{I}_n  \right ]\right ),
\end{equation}
where the covariance matrix $\mathbf{R}$ is expressed by
\begin{equation}\label{eq:RCF_1}
    \mathbf{R}=\rho _{f} \hat{\mathbf{G}}^{T}\mathbf{P}\mathbf{P}^{H}\hat{\mathbf{G}}^{\ast }\left ( \rho _{f}\tilde{\mathbf{G}}^{T}\mathbf{P}\mathbf{P}^{H}\tilde{\mathbf{G}}^{\ast } +\sigma _{w}^{2}\mathbf{I}_n\right )^{-1}
\end{equation}
In the clustered CF network, after scheduling $n_c \leq M_c$ out of $K_c$ UEs,
the received signal at cluster $c$ is
\begin{equation} \label{yc}
\begin{split}
\mathbf{y}_{c} & = \sqrt{\rho _{f}}\hat{\mathbf{G}}_{cc}^T\mathbf{P}_{c}\mathbf{x}_{c}+\sqrt{\rho _{f}}\tilde{\mathbf{G}}_{cc}^T\mathbf{P}_{c}\mathbf{x}_{c} \\
 & + \sum_{i=1,i\neq c}^{C}\sqrt{\rho _{f}}\hat{\mathbf{G}}_{ic}^T\mathbf{P}_{i}\mathbf{x}_{i}+\sum_{i=1,i\neq c}^{C}\sqrt{\rho _{f}}\tilde{\mathbf{G}}_{ic}^T\mathbf{P}_{i}\mathbf{x}_{i}+\mathbf{w}_{c}
\end{split}
\end{equation}
where $\mathbf{G}_{ic}=\hat{\mathbf{G}}_{ic}+\tilde{\mathbf{G}}_{ic}$ is ${M_i\times n_c}$ channel from APs of the cluster $i$ to the UEs of the cluster $c$, $\mathbf{P}_{i}$ is ${M_i\times n_c}$ linear precoding matrix, $\mathbf{x}_i=\left [ x_{i1},\ldots ,x_{in_c} \right ]^{T}$, $\mathbf{x}_i\sim \mathcal{CN}\left ( \mathbf{0},\mathbf{I}_{n_{c}} \right )$ is the symbol vector of the cluster $i$, $i\in \left \{ 1, 2, \ldots ,C \right \}$, and $\mathbf{w}_c=\left [ w_{c_{1}}\ldots,w_{c_{n_c}}  \right ]^{T}$ is the additive noise vector with $\mathbf{w}_c\sim \mathcal{CN}\left ( 0,\sigma_{w}^{2}\mathbf{I}_{n_c} \right )$. Therefore, the sum-rate of the cluster $c$ in this network is given by
\vspace{-1mm}
\begin{equation}\label{eq:RCL0}
    R_{c}=\log_{2}\left (  \det\left [\left ( \rho _{f}\hat{\mathbf{G}}_{cc}^T\mathbf{P}_{c}\mathbf{P}_{c}^{H}\hat{\mathbf{G}}_{cc}^* \right )\mathbf{R}_{c}^{-1}+\mathbf{I}_{n_c}  \right ]\right )
\end{equation}
and the covariance matrix $\mathbf{R}_{c}$ is described by
\vspace{-1mm}
\begin{equation} \label{eq:RCL1}
\begin{split}
\mathbf{R}_{c} & =E\left [ \left ( \mathbf{y}_{c}-\sqrt{\rho _{f}}\hat{\mathbf{G}}_{cc}^T\mathbf{P}_{c}\mathbf{x}_{c} \right )\left ( \mathbf{y}_{c}-\sqrt{\rho _{f}}\hat{\mathbf{G}}_{cc}^T\mathbf{P}_{c}\mathbf{x}_{c} \right )^{H} \right ] \\
 & =\rho _{f}\tilde{\mathbf{G}}_{cc}^T\mathbf{P}_{c}\mathbf{P}_{c}^{H}\tilde{\mathbf{G}}_{cc}^*+\sum_{i=1,i\neq c}^{C}\rho _{f}\hat{\mathbf{G}}_{ic}^T\mathbf{P}_{i}\mathbf{P}_{i}^{H}\hat{\mathbf{G}}_{ic}^*\\
 & +\sum_{i=1,i\neq c}^{C}\rho _{f}\tilde{\mathbf{G}}_{ic}^T\mathbf{P}_{i}\mathbf{P}_{i}^{H}\tilde{\mathbf{G}}_{ic}^*+\sigma _{w}^{2}\mathbf{I}_{n_c}
\end{split}
\end{equation}

where $\mathbf{x}_c$ and $\mathbf{w}_c$ are statistically independent. Finally, sum-rate of the total clustered network is given by
\begin{equation}\label{eq:RCL}
    R_{cl}=\sum_{c=1}^{C}R_{c}.
\end{equation}

\section{Proposed Sequential Multiuser Scheduling and Power Allocation}
\label{sec:US-PA}

\begin{figure}[htb]
  \centering
  \centerline{\includegraphics[width=8.5cm]{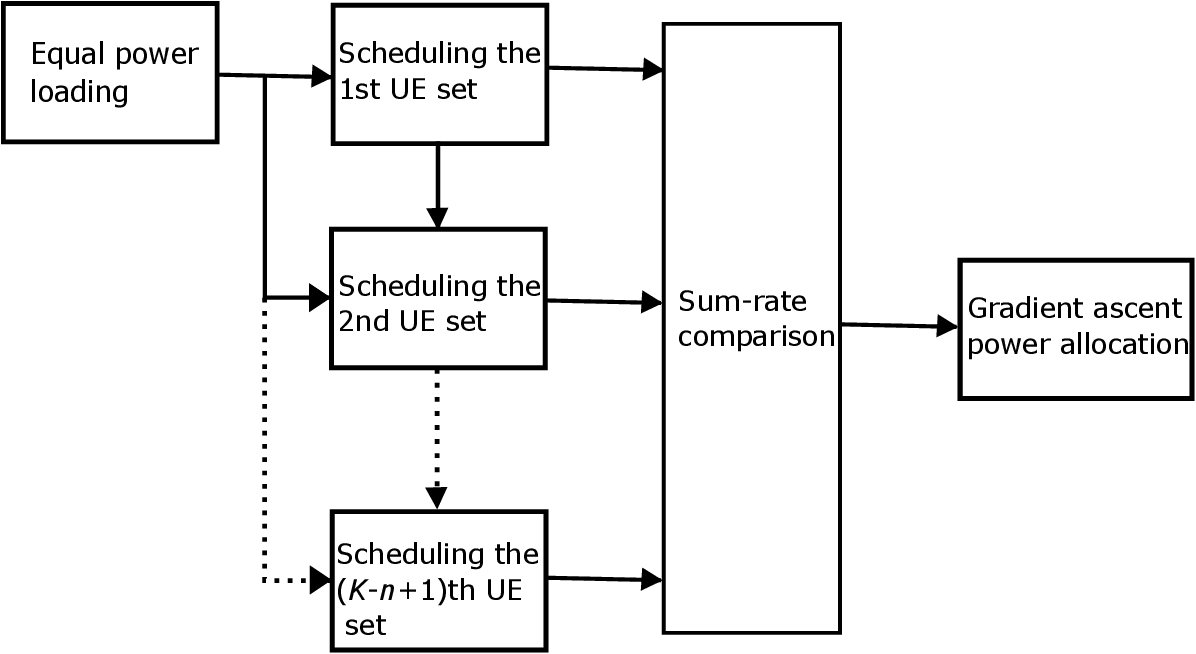}}
\caption{Block diagram of the proposed SMSPA resource allocation.}
\label{fig:fig1}
\end{figure}

In this section, we detail the proposed SMSPA scheme for multiuser scheduling and power allocation in clustered CF networks, which is outlined in Fig.~\ref{fig:fig1}. In particular, the SMSPA scheme employs an enhanced greedy algorithm for multiuser scheduling {using the method presented in \cite{mashdour2022enhanced},} in conjunction with a power allocation algorithm based on the GA method to maximize the sum-rate of the network. 
Unlike the enhanced greedy algorithm of the reference \cite{mashdour2022multiuser} which used equal power loading, we perform both scheduling and power allocation. Specifically, we first consider equal power loading and then employ the proposed greedy multiuser scheduling to schedule the best set using the sum-rate criterion and after that GA power allocation algorithm is employed. 

\subsection{Proposed ESG multiuser scheduling algorithm}
\label{ssec:Proposed}
In cluster $c$ including $\mathcal{U}_{c}=\left \{ 1,\ldots ,K_c \right \}$ as UEs and $M_c$ number of APs ($K_c>M_c$), in order to schedule a specific number of UEs such as $n_c$ so that $n_c\leq M_{c}$, we first adapt a greedy method similar to the approach applied in \cite{dimic2005downlink} to select the first set of UEs. However, unlike \cite{dimic2005downlink} we apply the MMSE precoder instead of the ZF precoder so that we can obtain a better performance and refine the search algorithm. In addition, the number of selected UEs is prespecified. For the selected set $S_{n_c}$ which results in a row-reduced channel matrix $\mathbf{G}_{cc}\left ( S_{n_c} \right )$, we aim at obtaining the solution to the optimization problem
\begin{equation}
\begin{aligned}
& \underset{S_{n_c}}{\text{max}}~R_{MMSE}\left ( S_{n_c} \right ) \\
& \text{subject to} \ \left \| \mathbf{P}_c\left ( S_{n_c} \right ) \right \|_{F}^{2}\leq P.
\end{aligned}
\end{equation}
where $R_{MMSE}\left ( S_{n_c} \right )$ is defined as the sum-rate with the MMSE precoder when $S_{n_c}$ is the set of intended users, $P$ is upper limit {to the covariance matrix of the received signal}, $\textup{trace}\left [ \mathbf{C}_{\mathbf{x}}\right ]\leq P$, and $\mathbf{P}_c\left ( S_{n_c} \right )=\mathbf{W}_{c}\mathbf{D}_{c}$ is the precoding matrix including the {normalized MMSE weight matrix} $\mathbf{W}_{c}\in\mathbb{C}^{M_c \times n_c} $ and the power {allocation matrix} $\mathbf{D}_{c}$ defined as
\begin{multline}
\mathbf{D}_{c}=\begin{bmatrix}
\sqrt{p_{1}} & 0 & \cdots  & 0\\
 0& \sqrt{p_{2}} & \cdots &0 \\
 \vdots & \vdots  &\cdots   & \vdots \\
 0&0  &\cdots   & \sqrt{p_{n_c} }
\end{bmatrix}=\textup{diag}\left ( \mathbf{d}_c \right )\\
, \mathbf{d}_c
=\left [ \sqrt{p_{1}} \ \sqrt{p_{2}} \ \cdots \  \sqrt{p_{n_c}} \right ]^T.
\end{multline}
We consider GA power allocation as described in Section \ref{ssec:Gradient} and the first set of UEs $S_{{n_c}\left ( 1 \right )}$ is obtained using the first stage of  Algorithm \ref{alg:alg1}. In order to assess more sets of the users so that we can approach the optimal set achievable by the exhaustive search method, we consider $S_{{n_c}\left ( 2 \right )}$ as the second set of UEs which is different from the set $S_{{n_c}\left ( 1 \right )}$ in only one UE. To obtain $S_{{n_c}\left ( 2 \right )}$, we remove the UE which has the lowest channel power among the UEs of the set $S_{{n_c}\left ( 1 \right )}$ shown by $u_{r}$ and replace it with the UE which possesses the most power among the UEs other than the set $S_{{n_c}\left ( 1 \right )}$ shown by $u_{a}$. Thus, we can show the removed and added UEs as follows, respectively,
\begin{equation} \label{kex}
         u_{r\left (1\right )}=\underset{u\in S_{{n_c}\left ( 1 \right )}}{{\arg \min}}~ \mathbf{g}_{u}^{H}\mathbf{g}_{u}
\end{equation}
\begin{equation} \label{knew}
 u_{a\left (1\right )}=\underset{u\in \mathcal{U}_{re\left ( 1\right )}}{{\arg \max}}~\mathbf{g}_{u}^{H}\mathbf{g}_{u}
\end{equation}
where $\mathbf{g}_{u}=\left [ g_{_{1,u}} \ \cdots \ g_{_{M_{c},u}} \right ]^{T}$ is the channel vector to UE $u$, and $\mathcal{U}_{re\left ( 1\right )}=\mathcal{U}_{c}\setminus S_{{n_c}\left ( 1 \right )}$ shows the remaining UE set other than $S_{{n_c}\left ( 1 \right )}$. The same process is done for the second set of UEs to find the third set. We continue to find new sets until we obtain $K_c-n_c$ sets of UEs beside the first set. We obtain the $i$th selected set of UEs and the $i$th remaining set of UEs as follows, respectively,
\begin{equation}
    S_{{n_c}\left ( i \right )}=\left (S_{{n_c}\left ( i-1 \right )}\setminus u_{r\left (i-1\right )} \right )\cup u_{a\left (i-1\right )}
\end{equation}
\begin{equation}
    \mathcal{U}_{re\left ( i\right )}=\mathcal{U}_{re\left ( i-1\right )}\setminus u_{a\left (i-1\right )}
\end{equation}
where  $i \in \left \{ 2,\cdots ,{K_{c}-n_c}+1 \right \}$. Finally, the desired set $S_{{n_c}_{d}}$ among the obtained sets, is the set which results in the highest sum-rate $R_{c}\left ( S_{{n_c}\left ( {J} \right )}\right )$, $J \in \left \{ 1,\cdots ,{K_{c}-n_c}+1 \right \}$ as derived in (\ref{eq:RCL0}).

\begin{algorithm}
\LinesNumbered
\SetKwBlock{Begin}{}{}
\caption{Proposed C-ESG Scheduling Algorithm.}\label{alg:alg1}
\SetAlgoLined
{{j=1}} \ \% first stage
  \Begin{
  \textbf{set} $l = 1$\;
  \textbf{find a user such that}

  $u_{1}=\underset{u\in \mathcal{U}_{c}}{\textup{argmax}} ~\mathbf{g}_{u}^{H}\mathbf{g}_{u}$\;
  \textbf{set $U_1 = {u_1}$ and denote the achieved rate}

  $R_{MMSE}\left ( U_{1} \right )$;

  }
\end{algorithm}

\begin{algorithm}
\LinesNumbered
\setcounter{AlgoLine}{6}
\SetKwBlock{Begin}{}{}
\SetAlgoLined
\nonl
  \Begin{
  \While  {$l<n_{c}$} {
   \hspace{0.5cm} $l=l+1$;

  \hspace{0.5cm} \textbf{find a user $u_{l}$ such that}

  \hspace{0.5cm} $u_{l}=\underset{u\in\left (  \mathcal{U}_{c} \setminus  U_{l-1} \right )}{\textup{argmax}}R_{MMSE}\left ( U_{l-1}\cup \left \{ u \right \} \right )$;

  \hspace{0.5cm} \textbf{set $U_{l}=U_{l-1}\cup \left \{ u_{l} \right \}$ and denote the rate}

  \hspace{0.5cm} $R_{MMSE}\left ( U_{l} \right )$;

  \hspace{0.5cm} \textbf{If $R_{MMSE}\left ( U_{l} \right )\leq R_{MMSE}\left ( U_{l-1} \right )$, break}

  \hspace{0.5cm} $l=l-1$;}
  $S_{n_{c}\left ( \textup{j} \right )}=U_{l}$\;
  \textbf{compute}: $ R_{c}\left ( S_{n_{c}\left ( \textup{j} \right )}\right )$\;
  $\mathcal{U}_{re\left ( \textup{j}\right )}=\mathcal{U}_{c}\setminus S_{{n_c}\left ( \textup{j} \right )}$\;
  $ u_{r\left (\textup{j}\right )}=\underset{u\in S_{{n_c}\left ( \textup{j} \right )}}{\textup{argmin}}\mathbf{g}_{u}^{H}\mathbf{g}_{u}$\;
  $u_{a\left (\textup{j}\right )}=\underset{u\in \mathcal{U}_{re\left ( \textup{j}\right )}}{\textup{argmax}}\mathbf{g}_{u}^{H}\mathbf{g}_{u}$\;

  }
  \For {$\textup{j}=2$ \textup{to}
{${K_{c}-n_{c}}+1$}} {
$S_{{n_c}\left ( \textup{j} \right )}=\left (S_{{n_c}\left ( \textup{j}-1 \right )}\setminus u_{r\left (\textup{j}-1\right )}  \right )\cup u_{a\left (\textup{j}-1\right )}$\;
$\mathcal{U}_{re\left ( \textup{j}\right )}=\mathcal{U}_{re\left ( \textup{j}-1\right )}\setminus u_{a\left (\textup{j}-1\right )}$\;
$ u_{r\left (\textup{j}\right )}=\underset{u\in S_{{n_c}\left ( \textup{j} \right )}}{\textup{argmin}}\mathbf{g}_{u}^{H}\mathbf{g}_{u}$\;
$u_{a\left (\textup{j}\right )}=\underset{u\in \mathcal{U}_{re\left ( \textup{j}\right )}}{\textup{argmax}}\mathbf{g}_{u}^{H}\mathbf{g}_{u}$\;
\textbf{compute}: $R_{c}\left ( S_{n_{c}\left ( \textup{j} \right )} \right )$;\
  }
$S_{{n_c}_{d}}=\underset{S_{{n_c}} \in S_{{n_c}\left ( {J} \right )}}{\textup{argmax}}\left \{  R_{c}\left ( S_{{n_c}} \right ) \right \}$\;
\textbf{Linear MMSE precoding of $n_{c}$ scheduled users}
\end{algorithm}

\subsection{{GA} power allocation algorithm}
\label{ssec:Gradient}

In Equation (\ref{yc}), the first part of the right hand side is the desired signal and remaining parts are the terms associated with imperfect CSI, inter-cluster interference and the noise. Therefore, by applying the power allocation, we  rewrite the estimated received signal at the cluster $c$ as follows
\begin{equation} \label{yc2}
\begin{split}
\mathbf{y}_{c} & = \sqrt{\rho _{f}}\hat{\mathbf{G}}_{cc}^T\mathbf{W}_{c}\mathbf{D}_{c}\mathbf{x}_{c}+\sqrt{\rho _{f}}\tilde{\mathbf{G}}_{cc}^T\mathbf{P}_{c}\mathbf{x}_{c} \\
 & + \sum_{i=1,i\neq c}^{C}\sqrt{\rho _{f}}\mathbf{G}_{ic}^T\mathbf{P}_{i}\mathbf{x}_{i}+\mathbf{w}_{c}
\end{split}
\end{equation}
For sum-rate maximization, we combine the received signal of the UEs using a linear receiver $\mathbf{a}=\frac{1}{\sqrt{K_c}}\mathbf{1}_{K_c}^{T}$, where $\mathbf{1}_{K_c}$ is a $1\times K_c$ vector of all 1 entities so that $\mathbf{a}^{H}\mathbf{a}=1$ and we obtain a simpler expression for sum-rate \cite{wang2012joint}. {After finding the power loading factors using the simplified sum-rate expression, we apply the obtained power allocation matrix in the sum-rate expressions defined in equations (\ref{eq:RCF}) and (\ref{eq:RCL0}) to determine the sum-rates.}

Thus, the combined received signal and the power ratio of the desired part of the signal to the interference and noise (SINR) are given as follows, respectively,
\begin{equation} \label{yc3}
\begin{split}
\mathbf{a}^T\mathbf{y}_{c} & = \sqrt{\rho _{f}}\mathbf{a}^T\hat{\mathbf{G}}_{cc}^T\mathbf{W}_{c}\mathbf{D}_{c}\mathbf{x}_{c}+\sqrt{\rho _{f}}\mathbf{a}^T\tilde{\mathbf{G}}_{cc}^T\mathbf{P}_{c}\mathbf{x}_{c} \\
 & + \sum_{i=1,i\neq c}^{C}\sqrt{\rho _{f}}\mathbf{a}^T\mathbf{G}_{ic}^T\mathbf{P}_{i}\mathbf{x}_{i}+\mathbf{a}^T\mathbf{w}_{c}
\end{split}
\end{equation}

\begin{equation}
    \textup{SINR}=\frac{\rho_{f} }{\sigma _{w}^{2}}\frac{\mathbf{a}^T\hat{\mathbf{G}}_{cc}^{T}\mathbf{W}_{c}\mathbf{D}_{c}\mathbf{D}_{c}^{H}\mathbf{W}_{c}^{H}\hat{\mathbf{G}}_{cc}^{*}\mathbf{a}}{\mathbf{a}^T\mathbf{Z}\mathbf{a}}
\end{equation}
where
\begin{equation}
    \mathbf{Z}=\tilde{\mathbf{G}}_{cc}^{T}\mathbf{P}_{c}\mathbf{P}_{c}^{H}\tilde{\mathbf{G}}_{cc}^{*}+\sum_{i=1,i\neq c}^{C}\rho_{f}{\mathbf{G}}_{ic}^{T}\mathbf{P}_{i}\mathbf{P}_{i}^{H}{\mathbf{G}}_{ic}^{T}+\mathbf{I}
\end{equation}
Assuming Gaussian signaling, the rate expression is obtained by $\frac{1}{2}\log_{2}\left ( 1+\textup{SINR} \right )$. Accordingly, the sum-rate expression is given by
\begin{equation} \label{SR2}
    SR=\frac{1}{2}\log_2\left [ 1+\frac{\rho_{f} }{\sigma _{w}^{2}}\frac{\mathbf{a}^T\hat{\mathbf{G}}_{cc}^{T}\mathbf{W}_{c}\mathbf{D}_{c}\mathbf{D}_{c}^{H}\mathbf{W}_{c}^{H}\hat{\mathbf{G}}_{cc}^{*}\mathbf{a}}{\mathbf{a}^T\mathbf{Z}\mathbf{a}} \right ].
\end{equation}
Equation (\ref{SR2}) is similar to $\frac{1}{2}\log_{2}\left ( 1+bx \right )$ where $b=\frac{\rho_{f} }{\sigma _{w}^{2}\mathbf{a}^T\mathbf{Z}\mathbf{a}}$ and $x=\mathbf{a}^T\hat{\mathbf{G}}_{cc}^{T}\mathbf{W}_{c}\mathbf{D}_{c}\mathbf{D}_{c}^{H}\mathbf{W}_{c}^{H}\hat{\mathbf{G}}_{cc}^{*}\mathbf{a}$ which is a monotonically increasing function of $x$, $b> 0$. Thus, we can maximize $x$ which is equivalent to the sum-rate using the following problem
\begin{equation} \label{opt.2.0}
\begin{aligned}
& \underset{\mathbf{d}_{c}}{\text{max}}~\left ( \mathbf{a}^T\hat{\mathbf{G}}_{cc}^{T}\mathbf{W}_{c}\textup{diag}\left ( \mathbf{d}_c \right )\textup{diag}\left ( \mathbf{d}_c \right )^{H}\mathbf{W}_{c}^{H}\hat{\mathbf{G}}_{cc}^{*}\mathbf{a}  \right ) \\
& \text{subject to} \left \| \mathbf{W}_c \textup{diag}\left ( \mathbf{d}_{c} \right ) \right \|^{2}\leq P.
\end{aligned}
\end{equation}
Since the objective function $x$ is scalar, $\textup{trace}(x)=x$. Therefore, by taking the derivative of the objective function with respect to the power loading matrix $\mathbf{D}_{c}$ and using the equality $\frac{\partial \textup{trace}\left ( \mathbf{AB} \right )}{\partial \mathbf{A}}=\mathbf{B}\odot \mathbf{I}$ where $\mathbf{A}$ is a diagonal matrix and $\odot$ shows the Hadamard product, we obtain
\begin{equation} \label{derx}
   \frac{\partial {x}}{\partial \mathbf{D}_{c}}=2\left ( \mathbf{W}_{c}^{H}\hat{\mathbf{G}}_{cc}^{*}\mathbf{a}\mathbf{a}^{T}\hat{\mathbf{G}}_{cc}^{T}\mathbf{W}_{c}\textup{diag}\left ( \mathbf{d}_{c} \right ) \right )\odot \mathbf{I}
\end{equation}
We can use a stochastic {GA} approach to update the power allocation coefficients as follows
\begin{equation} \label{eq1}
\begin{split}
\mathbf{d}_{c}\left ( i \right ) & = \mathbf{d}_{c}\left ( i-1 \right )+\lambda \frac{\partial {x}}{\partial \mathbf{D}_{c}} = \mathbf{d}_{c}\left ( i-1 \right )\\
 & +2\lambda \left ( \mathbf{W}_{c}^{H}\hat{\mathbf{G}}_{cc}^{*}\mathbf{a}\mathbf{a}^{T}\hat{\mathbf{G}}_{cc}^{T}\mathbf{W}_{c}\textup{diag}\left ( \mathbf{d}_{c}\left ( i-1 \right ) \right ) \right )\odot \mathbf{I}
\end{split}
\end{equation}
where $i$ and $\lambda$ represent the iteration index and the positive step size, respectively. 
 {Before running the adaptive algorithm, the transmit power constraint should be satisfied so that $\left \| \mathbf{W}_c \textup{diag}\left ( \mathbf{d}_{c} \right ) \right \|_{F}^{2}=\left \| \mathbf{P}_{c} \right \|_{F}^{2}\leq P$.} Therefore, the power scaling factor 
 {$\eta =\sqrt{\frac{\textup{trace}\left ( \mathbf{P}_{c}\mathbf{P}_{c}^{H} \right )}{\textup{trace}\left ( \mathbf{W}_{c}\textup{diag}\left ( \mathbf{d}_{c}.\mathbf{d}_{c} \right )\mathbf{W}_{c}^{H} \right )}}$}  is employed in each iteration to scale the coefficients properly. The adaptive power allocation is summarized in Algorithm \ref{alg:alg2} where $\textup{I}_{t}$ iterations are used. At the receiver, several detection and decoding techniques can be adopted \cite{jidf,spa,mfsic,mbsic,dfcc,jiomimo,jiocdma,mbdf,bfidd,1bitidd,locsme,okspme,lrcc,smce,wlmwf,vfap,did,rrmser,dynovs,aaidd,iddmtc,detmtc,1bitce,msgamp,msgamp2,comp}.

 \begin{algorithm}
\LinesNumbered
\SetKwBlock{Begin}{}{} \caption{{GA} Power Allocation Algorithm for
Sum-Rate Maximization.}\label{alg:alg2} \SetAlgoLined
\textbf{Input}\ $\mathbf{G}_{cc}$, $\mathbf{P}_{c}$,
$\mathbf{W}_{c}$, $\ \mathbf{a}$ \ \textup{and} \  $\lambda$\;

$\mathbf{d}_{c}\left ( 1 \right )=\mathbf{0}$\;

\For {${i}=2$ \textup{to}
{$\textup{I}_{t}$}} {
$\frac{\partial {x}}{\partial \mathbf{D}_{c}}=2\left ( \mathbf{W}_{c}^{H}\hat{\mathbf{G}}_{cc}^{*}\mathbf{a}\mathbf{a}^{T}\hat{\mathbf{G}}_{cc}^{T}\mathbf{W}_{c}\textup{diag}\left ( \mathbf{d}_{c}\left ( i-1 \right ) \right ) \right )\odot \mathbf{I}$\;
$\mathbf{d}_{c}\left ( i \right )=\mathbf{d}_{c}\left ( {i-1} \right )+\lambda \frac{\partial {x}\left ( \varepsilon  \right ) }{\partial \mathbf{D}_{c}}$\;

\If  {$\textup{trace}\left ( \mathbf{W}_{c}\textup{diag}\left ( \mathbf{d}_{c}\left ( i \right ).\mathbf{d}_{c}\left ( i \right ) \right )\mathbf{W}_{c}^{H} \right ) \neq \textup{trace}\left ( \mathbf{P}_{c}\mathbf{P}_{c}^{H} \right )$} {
   $\eta =\sqrt{\frac{\textup{trace}\left ( \mathbf{P}_{c}\mathbf{P}_{c}^{H} \right )}{\textup{trace}\left ( \mathbf{W}_{c}\textup{diag}\left ( \mathbf{d}_{c}\left ( i \right ).\mathbf{d}_{c}\left ( i \right ) \right )\mathbf{W}_{c}^{H} \right )}}$;

  $\mathbf{d}_{c}\left ( {i} \right )=\eta \mathbf{d}_{c}\left ( {i} \right )$;

  }

  }

\end{algorithm}

\section{SIMULATIONS}
\label{sec:simul}

In order to assess the proposed SMSPA resource allocation scheme, we
compare the sum-rate of the networks that use the proposed ESG,
standard greedy (SG), exhaustive search (ES) or WSR user scheduling
techniques and the proposed GA power allocation or equal power
loading (EPL). Note that we have adapted the WSR technique proposed
in \cite{ammar2021downlink} to the clustering method we have
implemented so that the WSR of the UEs in each cluster supported by
the corresponding APs is maximized. The considered CF network is a
squared area with the side length of 400~m equipped with $M$
randomly located APs and $K$ uniformly distributed UEs. Applying
network-centric clustering, we have considered $C$ = 4
non-overlapping clusters, where cluster c includes $M_c$ randomly
located APs and $K_c$ uniformly distributed UEs and the power
allocation is performed using {GA} algorithm. The large scale
coefficient in CF channel coefficient is modeled as $\beta
_{m,k}=\textup{PL}_{m,k} \times 10^{\frac{\sigma _{sh}z_{m,k}}{10}}$
where $10^{\frac{\sigma _{sh}z_{m,k}}{10}}$ is the shadow fading
with $\sigma _{sh}=8\textup{dB}$, $z_{m,k}\sim\mathcal{N}\left ( 0,1
\right )$, and $\textup{PL}_{m,k}$ is the path loss modeled as
\cite{tang2001mobile} \vspace{-2.5mm}
\begin{equation}
    \textup{PL}_{m,k}=\left\{\begin{matrix}
-\textup{D}-35\log_{10}\left ( d_{m,k} \right ), \textup{if }  d_{m,k}>d_{1}& \\
 -\textup{D}-10\log_{10}\left ( d_{1}^{1.5} d_{m,k}^2\right ), \textup{if }  d_{0}<d_{m,k}\leq d_{1}& \\
 -\textup{D}-10\log_{10}\left ( d_{1}^{1.5} d_{0}^2 \right ), \textup{if }  d_{m,k}\leq d_{0}&
\end{matrix}\right.
\end{equation}
where $d_{m,k}$ is the distance between the $m$th AP and
$k$th UE and $\textup{D}$ is
\vspace{-4mm}
\begin{multline}
\textup{D}=46.3+33.9\log_{10}\left ( f \right )-13.82\log_{10}\left ( h_{AP} \right )\\
-\left [ 1.11\log_{10}\left ( f \right )-0.7 \right ]h_{u}+1.56\log_{10}\left ( f \right )-0.8
\end{multline}
{where $f=1900$MHz is the carrier frequency, $h_{AP}=$15m,
$h_{u}=$1.5m are the AP and UE antenna heights, respectively,
$d_{0}=10$m and $d_{1}=$50m. If $d_{m,k}\leq d_{1}$ there is no
shadowing.}

\begin{figure}[htb]
  \centering
  \centerline{\includegraphics[width=8.5cm]{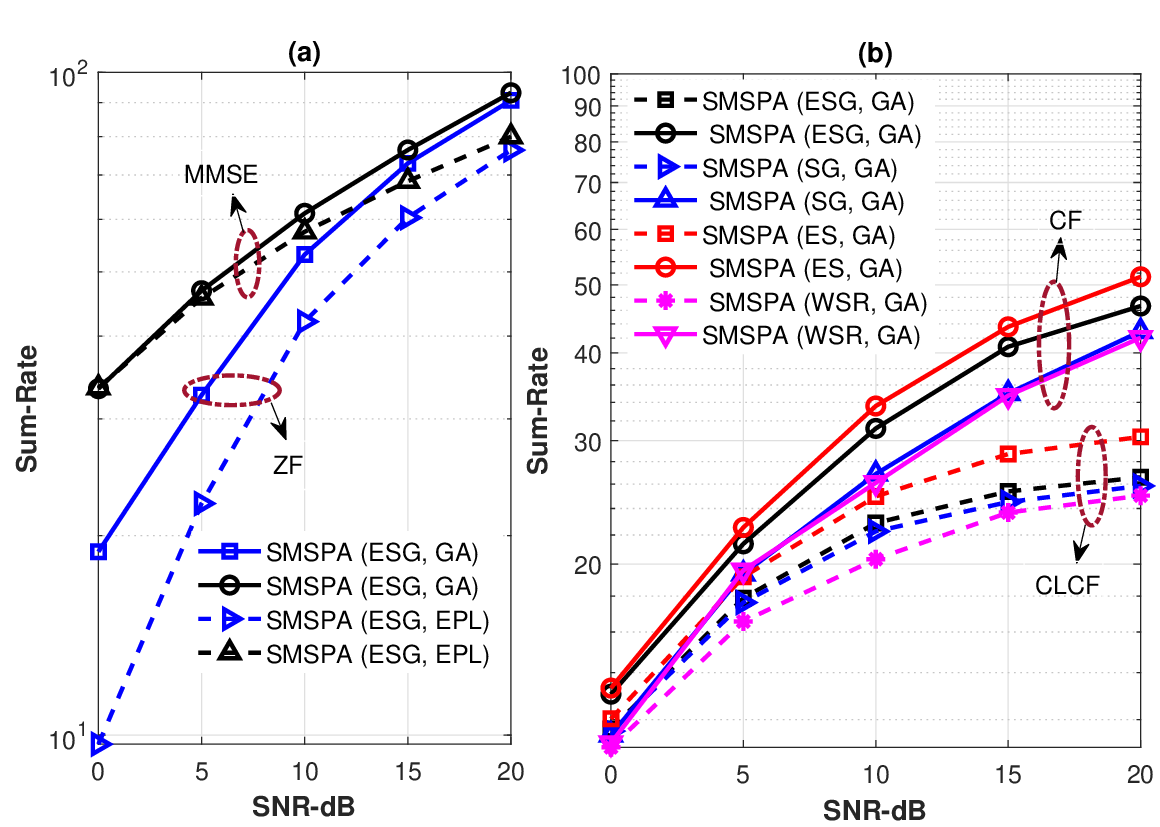}}
  \vspace{-2mm}
\caption{Performance of resource allocation schemes, (a): Comparison of the proposed SMSPA technique in CF networks for ZF and MMSE precoders with the system which has implemented the proposed ESG algorithm and equal power loading (EPL) ($M=64$, $K=128$, $n=24$), (b): Comparison of the different resource allocation techniques in CF and CLCF networks adapting GA power allocation when MMSE precoder is used ($M=64$, $K=16$, $n=8$).}
\label{fig:fig2}
\end{figure}

In Fig.~\ref{fig:fig2}(a), the sum-rate performances of the proposed
SMSPA scheme is assessed with the ESG scheduling algorithm using EPL
or the GA power allocation when ZF or MMSE precoders are applied.
While the sum-rates are increasing with the increase in the
signal-to-noise ratio (SNR), the MMSE precoder outperforms the ZF
precoder. In addition, the GA power allocation yields significant
performance improvement at low-to-medium SNR values.

Fig.~\ref{fig:fig2}(b) shows a comparison of different resource
allocation techniques when the MMSE precoder is used. We employ a
network with a small number of UEs while half of the UEs are
scheduled so that we can show the results for the ES method as well
as other methods. We notice that the proposed SMSPA resource
allocation which has used the ESG and GA algorithms has outperformed
other approaches and in the CF network the performance is close to
that of the optimal ES method. As expected and according to Equation
(\ref{yc}), CF shows better performance than that of the CLCF
network because of the extra interference terms caused by other
clusters. {We clarify that Fig.~\ref{fig:fig2} is plotted according
to the sum-rate expressions of equations (\ref{eq:RCF}) and
(\ref{eq:RCL0}) and simplified sum-rate equation of (\ref{SR2}) is
used only to derive the power loading factors.} However, as shown in
Table \ref{table:complexity}, the computational cost of the proposed
SMSPA scheme and the signaling load {as the number of channel
parameters} for CLCF network are substantially lower than CF.


%
%
%


  \vspace{-3mm}
\vspace{-2mm}
\begin{table}[htb!]
\caption{Computational complexity of the proposed SMSPA scheme in floating point operations and the signaling load in parameters for CF and CLCF networks when $M=64$, $K=128$ and $n=64$.}
\centering
\begin{tabular}{ |p{3cm}||p{2cm}|p{2cm}| }
 \hline
Network & CF &CLCF\\
 \hline
 Signaling load   & 24576    &6144\\
 \hline
 Computational cost &   $1.3632 \times 10^9$  & $69.354 \times 10^6$\\
 \hline
\end{tabular}

\label{table:complexity}
\end{table}

\section{CONCLUSIONS}
\label{sec:conc}
\vspace{-2mm}
This work has investigated resource allocation and sum-rate performance of the CF and the clustered CF networks with ZF and MMSE precoders. An SMSPA resource allocation scheme is developed that is based on  ESG multiuser scheduling and GA power allocation algorithms. Simulations have shown that the proposed SMSPA scheme has outperformed the existing methods and using the PA algorithm has also considerably improved the network performance compared with the EPL case. Additionally, in the case of the network clustering, a substantial computational complexity is saved using the proposed SMSPA scheme and the signaling load is much lower.

\bibliographystyle{IEEEbib}
\bibliography{refs}

\end{document}